\documentclass{iau}
\usepackage{graphicx}
\usepackage{amsmath}

\title[JD 11.~~The Needlet CMB Trispectrum] 
{The Needlet CMB Trispectrum}

\author[Antonino Troja, Simona Donzelli,Davide Maino \and Domenico Marinucci]   
{Antonino Troja$^{1,\dagger}$, Simona Donzelli$^2$,
 Davide Maino$^1$ \and Domenico Marinucci$^3$}

\affiliation{$^1$Dipartimento di Fisica, Universit\`a degli Studi di Milano, \\ Via Celoria 16, 20133, Milano (MI), Italy \\ $^\dagger$email: {\tt antonino.troja@unimi.it} \\[\affilskip]
$^2$INAF - Istituto di Astrofisica Spaziale e Fisica Cosmica, Milano, \\ Via E. Bassini 15, 20133, Milano (MI), Italy 
\\[\affilskip]
$^3$Dipartimento di Matematica, Universit\`a degli Studi di Roma Tor Vergata, \\ Via della Ricerca Scientifica, 00133, Roma (RM), Italy 
}

\pubyear{2014}
\volume{304}  
\pagerange{119--126}
\jname{Statistical Challenges in 21st Century Cosmology}
\editors{Cambridge University Press}
\begin{document}

\maketitle

\begin{abstract}
We propose a computationally feasible estimator for the needlet trispectrum, which develops earlier work on the bispectrum by \cite[Donzelli \etal\ (2012)]{Donzelli}. Our proposal seems to enjoy a number of useful properties, in particular a) the construction exploits the localization properties of the needlet system, and hence it automatically handles masked regions; b) the procedure incorporates a quadratic correction term to correct for the presence of instrumental noise and sky-cuts; c) it is possible to provide analytic results on its statistical properties, which can serve as a guidance for simulations. The needlet trispectrum we present here provides the natural building blocks for the efficient estimation of nonlinearity parameters on CMB data, and in particular for the third order constants $g_{NL}$ and $\tau_{NL}$.  
\end{abstract}

\firstsection 
\section{Introduction}
The Inflationary models describe the dynamics of the first instants of the Universe. Each model introduces a characteristic signature of non-Gaussianity into the anisotropy distribution of the Cosmic Microwave Background (CMB) radiation. The level of non-Gaussianity is described by the so-called \textit{Bardeen's Potential} (\cite[Bardeen (1980)]{11}):
\begin{equation}
	\Phi(\textbf{x}) = \Phi_L(\textbf{x}) + f_{NL}[\Phi_L^2(\textbf{x}) - \langle\Phi_L^2(\textbf{x})\rangle] + g_{NL}[\Phi_L^3(\textbf{x})]
\end{equation}
where $\Phi(\textbf{x})$ is the gravitational potential field of the Universe and $\Phi_L(\textbf{x})$ is its Gaussian part. The constants $f_{NL}$ and $g_{NL}$,
which parametrize the non-Gaussian part, can be measured from the amplitude of the bispectrum and trispectrum of the field, i.e. the harmonic counterpart of the 3-point and 4-point correlation function of the field, respectively. 

Although there exist several optimal bispectrum estimators that allow to evaluate $f_{NL}$ with high confidence (see for instance \cite[Komatsu \etal\ (2005)]{95a} or \cite[\textsc{Planck} Coll. (2013)]{95b}), the lack of an optimal trispectrum estimator has prevented the evaluation of strong constraints on $g_{NL}$ so far.


\section{Spherical Needlet System}
\subsection{Definition}
Spherical needlets (\cite[Narcowich \etal\ (2006)]{A}, \cite[Baldi \etal\ (2009)]{Baldietal}) are a wavelet system on the sphere defined by setting:
\begin{equation}
	\psi_{jk}(x):=\sqrt{\lambda_{jk}}\sum_lb\left(\frac{l}{B^j}\right)\sum_{m=-l}^lY_{lm}(\xi_{jk})\overline{Y_{lm}}(x);
\end{equation}
where $x\in S^2$, $\{\lambda_{jk},\xi_{jk}\}$ are a set of cubature points and weights on the sphere, $B>1$ is a constant related to the width of the needlet and $b(.)$ is a weight function satisfying the three following conditions: 
a) Compact Support, $b(\xi) > 0$ if $\xi\in(B^{-1},B)$, 0 otherwise; b) Partition of Unity, for all $\zeta\geq1$, $\sum_{j=0}^\infty b^2\left(\frac{\xi}{B^j}\right)=1$; c) Smoothness, $b(.)\in C^M$, i.e., $b(.)$ is $M$ times continuously differentiable, for some $M=1,2,\ldots$ or $M=\infty$.

\begin{figure}[!htbp]
\begin{center}
 \includegraphics[width=3.4in]{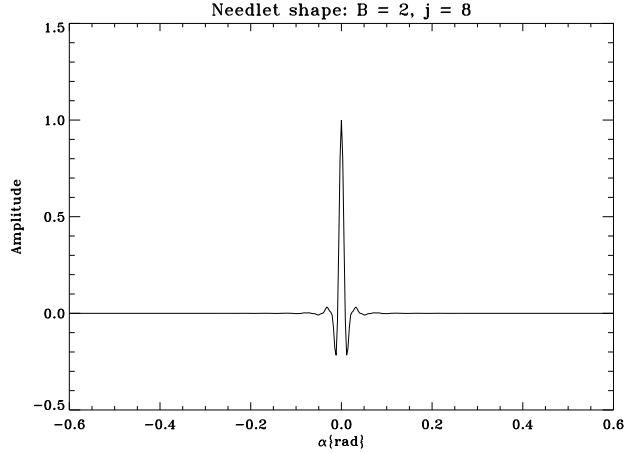} 
 \caption{1D-Spherical needlet. The figure shows the characteristic needle shape of the function, from which take its name (from \cite[Marinucci \etal\ (2008)]{se_stesso})}
   \label{fig:needlet_shape}
\end{center}
\end{figure}

\subsection{Reconstruction Formula}

The spherical needlet coefficients are provided by the analytical formula:
\begin{equation}\label{eq::needlet_coefficients}\beta_{jk} = \int_{S^2}T(x)\psi_{jk}(x)d\sigma(x)
		=\sqrt{\lambda_{jk}}\sum_lb\left(\frac{l}{B^j}\right)\sum_{m=-l}^la_{lm}Y_{lm}(\xi_{jk}).
\end{equation}

As a consequence of the partition of unity property, 
the following well-known reconstruction formula holds:
\begin{equation}
	f(x)=\sum_{jk}\beta_{jk}\psi_{jk}(x)\equiv\sum_{lm}a_{lm}Y_{lm}(x).
\end{equation}

\subsection{Localization and Uncorrelation Properties}

As argued earlier in \cite[Baldi \etal\ (2009)]{Baldietal}, spherical needlets enjoy an excellent localization property in real domain (fig. \ref{fig:needlet_shape}):
\begin{equation}\label{eq::localization_property}
	|\psi_{jk}(x)|\leq\frac{c_MB^j}{(1+B^j\arccos\langle\xi_{jk},x\rangle)^M}.
\end{equation}
More explicitly, spherical needlets are then quasi-exponentially localized around any cubature point $\xi_{jk}$.  Moreover, as extensively argued in the literature, the needlet coefficients evaluated on isotropic random fields are asymptotically uncorrelated, under mild regularity conditions. It is then possible to derive analytically their statistical properties, and to understand the role and expressions for correction terms under realistic experimental conditions (noise and masks, see Donzelli et al. for the bispectrum case) \\


\section{Optimal Trispectrum Estimator}
The sample trispectrum estimator is usually written using the spherical harmonics coefficients as:
\begin{equation}
	\sum_{m_1m_2m_3m_4} \mathcal {G}(l_1,m_1,l_2,m_2,l_3,m_3,l_4,m_4)\times a_{l_1m_1}a_{l_2m_2}a_{l_3m_3}a_{l_4m_4},
\end{equation}
where $\mathcal{G}(l_1,m_1,l_2,m_2,l_3,m_3,l_4,m_4)$ is the Gaunt integral defined for instance in \cite[Marinucci \and Peccati (2011)]{MarPec}.
This estimator is clearly unfeasible under realistic experimental conditions, due to the presence of missing data and anisotropic noise. As a consequence of the previous discussion, it is possible to exploit the needlet coefficients to derive an alternative, computationally feasible and statistically sound estimators of the trispectrum. More precisely, exploiting Wick theorem on higher order moments of Gaussian variables we propose the following needlet trispectrum:
\begin{equation}
\begin{split}\label{eq::needlet_trispectrum_estimator}
	J^{j_1j_2}_{j_3j_4} =&\frac1{\sigma_{j_1}\sigma_{j_2}\sigma_{j_3}\sigma_{j_4}}\sum_{k}[\beta_{j_1k}\beta_{j_2k}\beta_{j_3k}\beta_{j_4k} \\&- \{\langle\beta_{j_1k}\beta_{j_2k}\rangle\beta_{j_3k}\beta_{j_4k} + 5\,perms.\} + \{\langle\beta_{j_1k}\beta_{j_2k}\rangle\langle\beta_{j_3k}\beta_{j_4k}\rangle + 2\,perms.\}]\\
	\end{split}
\end{equation}

Heuristically, the needlet trispectrum is constructed combining quadruples of  coefficients, evaluated at the scales of interest, and subtracting linear and quadratic components in order to cancel bias and minimize the variance. Further details, and a software which exploits the needlet trispectrum to evaluate  $g_{NL}$ on CMB maps, will be provided in the forthcoming paper \cite[Troja \etal\  (2014) ]{Troja}. 

\end{document}